\documentclass[published]{JHEP3} 

\JHEP{00(0000)000}

\JHEPspecialurl{http://jhep.sissa.it/JOURNAL/JHEP3.tar.gz}

\usepackage{epsfig,multicol,bbm}

\newcommand\fverb{\setbox\pippobox=\hbox\bgroup\verb}
\newcommand\fverbdo{\egroup\medskip\noindent%
			\fbox{\unhbox\pippobox}\ }
\newcommand\fverbit{\egroup\item[\fbox{\unhbox\pippobox}]}
\newbox\pippobox

\def\gtap{\mathrel{ \rlap{\raise 0.511ex \hbox{$>$}}{\lower 0.511ex
   \hbox{$\sim$}}}} 
\def\ltap{\mathrel{ \rlap{\raise 0.511ex
    \hbox{$<$}}{\lower 0.511ex \hbox{$\sim$}}}} 
\newcommand{\bea}{\begin{eqnarray}} 
\newcommand{\eea}{\end{eqnarray}}
\def\beq{\begin{equation}}
\def\enq{\end{equation}}
\def\ba{\begin{eqnarray}}
\def\ea{\end{eqnarray}}
\newcommand{\bfB}{\bf B}
\newcommand{\g}{\gamma}
\newcommand{\w}{\omega}

\newcommand{\be}{\begin{equation}}
\newcommand{\ee}{\end{equation}}
\newcommand{\gag}{g_{a\gamma}}
\def\<{\langle}
\def\>{\rangle}

\title{Signatures of photon and axion-like particle mixing in the
  gamma-ray burst jet}

\author{ {Olga Mena,$^{a}$ Soebur Razzaque,$^{b,}$\thanks{Present
      address: Naval Research Laboratory, Washington, DC 20375, USA}
    ~and F. Villaescusa-Navarro$^{a}$} \\ {$^{a}$IFIC, Universidad de
    Valencia-CSIC, E-46071, Valencia, Spain} \\ {$^{b}$College of
    Science, George Mason University, Fairfax, Virginia 22030, USA}
  \\ E-mail: \email{omena@ific.uv.es},
  \email{srazzaqu@gmu.edu},
  \email{francisco.Villaescusa@ific.uv.es} }

\received{January 1, 0000} 		
\revised{January 1, 0000}
\accepted{January 1, 0000}		

\preprint{\hepth{9912999}}	

\abstract{Photons couple to Axion-Like Particles (ALPs) or more
  generally to any pseudo Nambu-Goldstone boson in the presence of an
  external electromagnetic field.  Mixing between photons and ALPs in
  the strong magnetic field of a Gamma-Ray Burst (GRB) jet during the
  prompt emission phase can leave observable imprints on the gamma-ray
  polarization and spectrum.  Mixing in the intergalactic medium is
  not expected to modify these signatures for ALP mass $> 10^{-14}$~eV
  and/or for $<$~nG magnetic field.  We show that the depletion of
  photons due to conversion to ALPs changes the linear degree of
  polarization from the values predicted by the synchrotron model of
  gamma ray emission.  We also show that when the magnetic field
  orientation in the propagation region is perpendicular to the field
  orientation in the production region, the observed synchrotron
  spectrum becomes steeper than the theoretical prediction and as
  detected in a sizable fraction of GRB sample. Detection of the
  correlated polarization and spectral signatures from these
  steep-spectrum GRBs by gamma-ray polarimeters can be a very powerful
  probe to discover ALPs.  Measurement of gamma-ray polarization from
  GRBs in general, with high statistics, can also be useful to search
  for ALPs.}


\begin{document}

\section{Introduction}

Axions represent the most convincing and elegant solution to the
strong CP problem~\cite{PecceiQuinn}.  They are the pseudo
Nambu-Goldstone bosons of a global $U(1)_{PQ}$
symmetry~\cite{weinberg,wilczek}.  Axion-Like Particles (ALPs) can be
understood as generalizations of the axions, and appear generally in
theories beyond the Standard Model (SM) of Particle Physics.  Axions
and ALPs couple to photons in the presence of an external
electromagnetic field \cite{raffeltstodolsky}.  For axions the
strength of the coupling $\gag$ is inversely proportional to the
energy scale $M$ at which the $U(1)_{PQ}$ symmetry is spontaneously
broken, and is directly related to the particle mass $m_a$.  For ALPs
there is no general relation between the energy scale of the new
physics beyond the SM and the ALP mass, and therefore in the following
we shall consider the parameters $\gag$ and $m_a$ to be independent of
each other.  ALPs may be copiously produced in the early universe,
either thermally~\cite{turner} or non-thermally~\cite{kolb}, providing
a possible (sub) dominant (hot) dark matter candidate.

Mixing of photons and ALPs in an electromagnetic field results in
photon-ALP conversion and a change in photon polarization states.  The
former effect has been extensively exploited to search for ALPs that
are created in the Sun, travel to the Earth as ALPs and convert to
$\sim$ keV photons in the magnetic field of a laboratory experiment.
From non-detection of such photons, the CAST experiment has reported a
lower bound on the ALP energy scale of $M>1.1\cdot 10^{10}$~GeV, which
translates to a constraint on the photon-ALP coupling of $\gag
<8.8\cdot 10^{-11}$~GeV$^{-1}$ for ALP masses of $m_a\sim
0.02$~eV~\cite{cast}.  These constraints exclude a region in the
$\gag$--$m_a$ parameter space.  The same conversion mechanism is used
by the ADMX experiment to search for ALP dark matter that converts to
microwave photons~\cite{admx}.

Observation of supernovae (SNe) Ia dimming has been suggested as a
possible signature of photon-to-ALP conversion, thus depleting the
photon flux, in the Inter-Galactic Magnetic Field
(IGMF)~\cite{csaki,deffayet,ostman,basset,raffelt,verde}.  This is an
alternative to the standard interpretation by a dark energy fluid that
is responsible for recent accelerated expansion of the Universe,
making the distances of the SNe Ia larger.  Search for circular
polarization in the Cosmic Microwave Background (CMB) data has been
proposed as probe of photon-ALP mixing in the
IGMF~\cite{Agarwal:2008ac}.  Mixing in the IGMF has also been
considered as a possible mechanism to produce ultra-high energy
cosmic-ray events, assumed to be photons which are not attenuated
while in their ALP states and while propagating from distant sources
to the Earth~\cite{kaloper}.  A similar mechanism has been proposed to
search for photon-ALP conversion effects in the GeV--TeV $\g$-ray
fluxes from distant active galactic
nuclei~\cite{raffelt2,prada,roncadelli,bassan2}.  Detection of these
fluxes at very high energies may provide hints of photon-ALP mixing,
which would be absorbed by the Extragalactic Background Light (EBL)
otherwise (see e.g.\ Ref.~\cite{Razzaque:2008te}).

Here we study the observational consequences of a photon-ALP coupling
on GRB photon polarization and fluxes at \emph{low energies}, in the
$\sim$ keV--MeV range, arising from high magnetic field in the GRB
jet.~\footnote{GRB pseudo-Goldstone boson emission and its subsequent
  conversion to electromagnetic energy was proposed as a possible
  mechanism for the observed GRBs~\cite{loeb,bertolami,bere}.  We do
  not address such a possibility in the present study.}  A search in
this energy range has several advantages: (i) GRBs, the most powerful
explosions in the Universe, release upwards of $10^{53}$ erg of
isotropic-equivalent $\g$-ray energy, mostly in the $\sim$ keV--MeV
range~\cite{Meegan:1992xg}.  Thus we have the most powerful photon
beam at our disposal to investigate the effect.  (ii) Unlike TeV $\g$
rays, MeV photons are not attenuated in the EBL and detection of any
effect due to photon-ALP coupling does not depend on the EBL models
(see e.g. Refs.~\cite{Stecker:2005qs, Kneiske:2003tx,
  Franceschini:2008tp, Gilmore:2009zb, Finke:2009xi}).  (iii) Since
photons are converted to ALPs in the high magnetic field of the GRB
jet, ALPs may not convert back to photons while propagating in the
IGMF and in the galactic magnetic fields\footnote{They can, however,
  convert back to photons in a suitable laboratory experiment.}~.
Search for photon-ALP mixing in the photon polarization data has been
suggested for GRBs in the past, both in the strong magnetic field of
the GRB~\cite{rubbia} and in the IGMF~\cite{bassan} (see also
Ref.~\cite{chameleon}).  We have studied effects of photon-ALP mixing
on GRB $\g$-ray polarization using a realistic emission model, namely
synchrotron radiation by relativistic electrons in the strong magnetic
field, either advected from the GRB central engine~\cite{Spruit:2001,
  Lyutikov:2003bz} or generated in the shocks~\cite{Rees:1994nw,
  Katz:1994, Tavani:1996, Medvedev:1999tu} or both.

In the synchrotron model, which is also the leading model for observed
$\sim$ keV--MeV $\g$-ray emission, a population of electrons are
assumed to be injected as a power-law above a minimum particle Lorentz
factor in the magnetized plasma with an optical depth less than unity.
The peak of the observed energy spectrum ($E^2\,dN/dE$), typically in
the $\sim$ 0.1--1~MeV range, is identified with the characteristic
synchrotron frequency from the electrons with the minimum Lorentz
factor in the comoving GRB jet frame, boosted by the bulk Lorentz
factor of the jet.  Synchrotron radiation is partially polarized with
a linear polarization degree of $\approx 50\%$ at frequencies much
lower than the characteristic frequency, reaching $\approx 70\%$ at
the maximum~\cite{Rybicki:1979}.  We model the initial polarization
states of the observed photons in the $\sim$ keV--MeV $\g$ rays as
from the electrons with minimum Lorentz factor, according to
synchrotron radiation theory in the comoving frame.  The effect of
photon-ALP mixing then changes the observed polarization from the
expected pattern.

To date, prompt $\g$ ray polarization has been measured from only a
handful of GRBs, most notably an $(80\pm 20)\%$ linear polarization
from GRB 021206 by {\em RHESSI}~\cite{detect}.  However, these
measurements are statistically inconclusive and suffer from large
systematic uncertainties (see e.g. Ref.~\cite{Rutledge:2003wa}).
Gamma-Ray Burst Polarimeter (GAP), sensitive in the 50--300~keV range,
aboard the recently launched {\it IKAROS Solar Sail} is one of the new
generation of instruments to measure $\g$ ray
polarization~\cite{Yonetoku:2010nq}.  A number of satellite missions
such as the Advanced Compton Telescope (ACT)~\cite{ACT}, Gamma-ray
Burst Investigation via Polarimetry and Spectroscopy
(GRIPS)~\cite{GRIPS}, and Polarimeters for Energetic Transients
(POET)~\cite{POET} are also being planned to measure $\g$ ray
polarization in the keV--MeV range.  These experiments are expected to
measure GRB polarization with a high statistical significance and have
been shown to be excellent tools to test the synchrotron emission
models (see e.g. Ref.~\cite{toma}).  Eventually, these broadband
polarimeters will be able to detect deviations from the standard
synchrotron polarization pattern.  Such frequency-dependent deviations
in the polarization pattern could be explained in terms of photon-ALP
mixing.

The polarization pattern induced by photon-ALP mixing can be
accompanied with a detectable change in the $\g$-ray spectral slope,
due to a depletion of preferentially low energy photons that convert
to ALPs in the GRB jet.  Indeed a specific prediction of the GRB
synchrotron model is that, below the peak energy the spectrum can not
be harder than the photon index $\alpha_\g = -2/3$, where $dN/dE
\propto E^{\alpha_\g}$, a limit that arises from synchrotron theory of
radiation from a single particle~\cite{Rybicki:1979,Jackson:1998}.
Observed variation of the GRB low-energy spectra softer than this
limit may be explained as cooling effect on the electron spectrum,
producing a $\g$-ray spectrum as soft as $dN/dE \propto E^{-3/2}$
(see, e.g., Ref.~\cite{Sari:1997qe}).  Majority of bright GRBs,
detected by the Burst And Transient Source Experiment (BATSE) aboard
the {\em Compton Gamma Ray Observatory}, for which good spectral data
are available~\cite{Kaneko:2006qe} falls within the synchrotron limit
of $-2/3 \ge \alpha_\g \ge -3/2$.  However a significant ($\sim$ 20\%)
fraction violates the ``synchrotron death line'' of $\alpha_\g = -2/3$
\cite{Preece:1998jy}, and a ``harder when brighter'' tendency is
present in the data.  The same effect has been detected in
time-integrated and time-resolved spectra from joint observations by
the Burst Alert Telescope (BAT) aboard {\em Swift} and by the Wide
band All-sky Monitor (WAM) aboard {\em Suzaku}~\cite{Krimm:2009nb},
and most recently by the Gamma-ray Burst Monitor (GBM) aboard the {\em
  Fermi Gamma-ray Space Telescope}~\cite{Ghirlanda:2010}.  We predict
that polarization measurements of these steep-spectrum GRBs can shed
light, or even lead to discovery of ALPs.

The structure of the paper is as follows.  We review the photon-ALP
mixing phenomena in Sec.\ 2 and apply this formalism to the GRB jet
and synchrotron emission model in Sec.\ 3.  We discuss our results in
Sec.\ 4 and conclude our study in Sec.\ 5.

\section{Photon-ALP mixing and conversion probabilities}

We follow here the photon-axion/ALP interaction formalism from
Ref.~\cite{raffeltstodolsky} (see also Ref.~\cite{bassan}). The
lagrangian for the photon-ALP system is given by\footnote{We adopt the
  natural unit convention $\hbar=c=1$.}
\ba
{\cal L} &=& -\frac{1}{4} F_{\mu \nu} \, F^{\mu \nu} + 
\frac{{\alpha}^2}{90 \, m^4_e}\, 
\left[ \left(F_{\mu \nu} \, F^{\mu \nu} \right)^2 + 
\frac{7}{4} \left(F_{\mu \nu}\, \tilde F^{\mu \nu} \right)^2 \right]
\nonumber \\ && +
\frac{1}{2} \, \partial^{\mu} a \, \partial_{\mu} a - 
\frac{1}{2} \, m_{a}^2 \, a^2 -
\frac{1}{4} \,\gag F_{\mu\nu}\tilde{F}^{\mu\nu} a~,
\label{eq:lag}
\ea
where $F_{\mu \nu}$ is the electromagnetic field tensor,
$\tilde{F}_{\mu\nu} =\frac{1}{2} \epsilon_{\mu\nu\rho\sigma}
F^{\rho\sigma}$ is its dual, $\alpha$ is the fine-structure constant
and $m_e$ is the electron mass. The second term in Eq.~(\ref{eq:lag})
is the Euler-Heisenberg effective Lagrangian, which accounts for
one-loop corrections to the classical electrodynamics. The third and
fourth terms in Eq.~(\ref{eq:lag}) are the Lagrangian terms describing
the ALP field $a$ with a mass $m_a$.  The last term is the
photon-ALP interaction lagrangian, which, in terms of the external
electromagnetic field, reads
\begin{equation}
{\cal L}_{a\gamma}=-\frac{1}{4} \,\gag
F_{\mu\nu}\tilde{F}^{\mu\nu}a=\gag \, {\bf E}\cdot{\bf B}\,a~.
\label{eq:int}
\end{equation}
Here $\gag$ is the photon-ALP coupling constant, {\bf E} and $\bfB$
are the electric and magnetic fields respectively.

The evolution equations for a mono-energetic photon/ALP beam with
energy $\omega$ propagating along the $z$ direction in an external and
homogeneous magnetic field transverse (${\bf B}_T$) to the beam
direction (i.e. in the $x$-$y$ plane) are given by:
\bea
\omega^2 A_\perp+\partial^2_z A_\perp +
\frac{4 \alpha}{45\pi}\left(\frac{B_T}{B_{\rm crit}}\right)^2 
\omega^2 A_\perp -\frac{4 \pi n_e\alpha}{m_e} A_{\perp}&=&0~,
\nonumber \\
\omega^2 A_\parallel+\partial^2_z A_\parallel +
\frac{7 \alpha}{45\pi}\left(\frac{B_T}{B_{\rm crit}}\right)^2 
\omega^2 A_\parallel +\omega \gag B_T a -
\frac{4 \pi n_e\alpha}{m_e} A_{\parallel}&=&0~,
\nonumber \\
\omega^2 a +\partial^2_z a -m^2_a a +
\omega \gag B_T A_{\parallel}&=&0~.
\label{eq:evol}
\eea
Here $A_\perp$ and $A_{\parallel}$ are the two photon polarization
components (both in the $x$-$y$ plane) perpendicular and parallel to
the external magnetic field ${\bf B}_T$, respectively.  The plasma
term in the equations of motion arises due to the presence of
electrons in the media, giving an effective mass to the photons, and
is proportional to the electron number density $n_e$.  The critical magnetic
field is defined as $B_{\rm crit}\equiv m^2_e/e = 4.414\cdot
10^{13}$~G, where $e$ is the electron charge.  In the limit where
$\omega \gg m_a$, the evolution of the system can be linearized
in the form of a first order differential equation\footnote{We follow
the notation adopted in Ref.~\cite{bassan}.}
\begin{equation}
\left(i \, \frac{d}{d z} + \omega +  {\cal M} \right)  
\left(\begin{array}{c}A_\perp (z) \\ A_\parallel (z) \\ a (z) 
\end{array}\right) = 0~.
\label{eq:mix}
\end{equation}
Here ${\cal M}$ is a mixing matrix of the axion field with the photon
polarization components, and is given by
\begin{equation}
{\cal M} =   \left(\begin{array}{ccc}
\Delta_{ \perp}  & 0 & 0 \\
0 &  \Delta_{ \parallel}  & \Delta_{a \gamma}  \\
0 & \Delta_{a \gamma} & \Delta_a 
\end{array}\right)~.
\label{eq:mix1}
\end{equation}
The elements of ${\cal M}$ can be expressed as $\Delta_\perp \equiv
2\Delta_{\rm QED} + \Delta_{\rm pl}$, $\Delta_\parallel \equiv
(7/2)\Delta_{\rm QED} + \Delta_{\rm pl}$ following Ref.~\cite{bassan},
and we provide their reference values relevant in our case below
\ba
\Delta_{\rm QED} &\equiv & \frac{\alpha \omega}{45\pi} 
\left( \frac{B_T}{B_{\rm cr}} \right)^2 
\simeq 1.34\cdot 10^{-12} \left( \frac{\omega}{{\rm keV}} \right)
\left( \frac{B_T}{10^6\,{\rm G}} \right)^2 ~{\rm cm}^{-1},
\nonumber \\ 
\Delta_{\rm pl} &\equiv & -\frac{\omega^2_{\rm pl}}{2\omega} \simeq
- 3.49\cdot 10^{-12} \left( \frac{\omega}{{\rm keV}} \right)^{-1}
\left( \frac{n_e}{10^8\,{\rm cm}^{-3}} \right) ~{\rm cm}^{-1},
\nonumber \\
\Delta_{a\g} &\equiv & \frac{1}{2}g_{a\g} B_T
\simeq 1.32\cdot 10^{-11}
\left( \frac{g_{a\g}}{8.8\cdot 10^{-11}\,{\rm GeV}^{-1}} \right)
\left( \frac{B_T}{10^6\,{\rm G}} \right) ~{\rm cm}^{-1},
\nonumber \\ 
\Delta_{a} &\equiv & -\frac{m_a^2}{2\omega}
\simeq -2.53\cdot 10^{-13}
\left( \frac{\omega}{{\rm keV}} \right)^{-1}
\left( \frac{m_a}{10^{-7}\,{\rm eV}} \right)^2 ~{\rm cm}^{-1} .
\label{matrix_elements}
\ea
The plasma frequency is defined as $\omega_{\rm pl} = \sqrt{4\pi\alpha
n_e/m_e} =3.71\cdot 10^{-14}\sqrt{n_e/{\rm cm}^{-3}}$~keV.  Notice
from Eqs.~(\ref{eq:mix}) and (\ref{eq:mix1}) that the component of the
photon beam polarization perpendicular to the ${\bf B}_T$ field,
$A_\perp$, will decouple from the evolution of the photon-ALP system.
In other words the ALP couples only to the $A_\parallel$ polarization
component.

A generalization of the scenario discussed so far is when ${\bf B}_T$
makes an angle $\xi$, $0\le\xi\le 2\pi$, with the $y$ axis in a fixed
coordinate system.  A rotation of the mixing matrix
[Eq.~(\ref{eq:mix1})] in the $x$-$y$ plane then leads to a new form
and the evolution equation of the photon-ALP system reads
\begin{equation} i\frac{d}{dz}
\left(\begin{array}{c}A_\perp (z) \\ A_\parallel (z) \\ a (z)
\end{array}\right)=- \left(\begin{array}{ccccccccc}
\Delta_\perp\cos^2\xi+\Delta_\parallel\sin^2 \xi & \cos\xi \sin\xi
(\Delta_\parallel-\Delta_\perp) & \Delta_{a \gamma}\sin \xi \\ \cos\xi
\sin\xi (\Delta_\parallel-\Delta_\perp) &
\Delta_\perp\sin^2\xi+\Delta_\parallel\cos^2 \xi & \Delta_{a
\gamma}\cos \xi \\ \Delta_{a \gamma}\sin \xi & \Delta_{a \gamma}\cos
\xi & \Delta_a \\ \end{array} \right) \left(\begin{array}{c} A_\perp
(z) \\ A_\parallel (z) \\ a (z) \end{array}\right)~.
\label{eq:evolxi} 
\end{equation} 
If there are more than one magnetic field domain present in the
problem, then Eq.~(\ref{eq:evolxi}) needs to be solved for each domain
with appropriate initial conditions.  Under the assumptions that all
the domains in a particular environment (constant $n_e$ and the same
initial conditions for the fields at $z=0$) have identical coherence
lengths and magnetic field strengths, and only the orientation of the
magnetic field ${\bf B}_T$ in each domain is random, then the average
effect can be calculated by randomly varying $\xi$.  We mainly
consider the scenario where photons are created at $z=0$, at source,
and cross ${\bf B}_T$ field domains where (i) $\xi=0$ or $\pi/2$ in
each domain, and (ii) $\xi$ is random.  In both cases each photon
crosses only one coherence length width in the $z$ direction.  The
final beam consists of contributions from all domains.  This is
different from propagation of the beam in the intergalactic medium
where each photon/ALP crosses many IGMF domains and the initial
conditions change each time the beam enters a domain (see,
e.g. Ref.~\cite{bassan}).

In analogy with two-family neutrino mixing, the conversion probability
of $A_\parallel$ into ALPs after traveling a coherence length $L$ and
for $\xi =0$ reads
\begin{equation}
P_{a \gamma} =\sin^2 2 \theta 
\sin^2 \left( \frac{\Delta_{\rm osc} \, L}{2} \right),
\label{eq:probosc}
\end{equation}
where the oscillation wave number is ${\Delta}_{\rm osc} =
\sqrt{(\Delta_a - \Delta_{\parallel})^2 + 4 \Delta_{a\g}^2}$ and the
mixing angle is $\theta = (1/2) \arctan [2 \Delta_{a
\gamma}/(\Delta_{\parallel}-\Delta_a)]$.  From Eq.~(\ref{eq:probosc})
it is possible to infer the energy range in which the conversion
probabilities are approximately energy independent and mixing effects
will be maximal ($\theta \approx \pi/4$) for $\omega_L \le \omega \le
\omega_H$, where the low and high critical energies, respectively, are
given by~\cite{roncadelli,bassan2,bassan}
\ba
\omega_L &\equiv& \frac{E\, |\Delta_a- \Delta_{\rm pl}|}
{2\, \Delta_{a \gamma}} \simeq  
\frac{0.12\, | m_a^2 - {\omega}_{\rm pl}^2|}{(10^{-7}{\rm eV})^2}
\left( \frac{B_T}{10^{6}~{\rm G}} \right)^{-1} 
\left( \frac{\gag}{8.8\cdot 10^{-11}~{\rm GeV}^{-1}} \right)^{-1} 
{\rm keV},~~~ \nonumber \\
\omega_H &\equiv& \frac{90\pi\,\gag\, B^2_{\rm cr}}{7\alpha\,B_T} 
\simeq 5.62
\left(\frac{B_T}{10^{6}~{\rm G}} \right)^{-1} 
\left(\frac{\gag}{8.8\cdot 10^{-11}~{\rm GeV}^{-1}} \right) 
{\rm keV}.
\label{LHenergy}
\ea 
We focus on the specific problem of photon-ALP mixing in the GRB jet
and the impact of the photon-ALP conversions in the observed photon
spectrum in the next section.

\section{Gamma-ray emission and conversion to ALPs in the GRB jet}

Synchrotron radiation from relativistic electrons that are accelerated
in the GRB jet, either due to dissipation of the jet kinetic energy
(internal shocks of plasma shells)~\cite{Rees:1994nw} or magnetic
flux~\cite{Spruit:2001} from a central engine, is believed to be the
dominant mechanism to produce observed $\g$ rays in the keV--MeV
range.  In the internal shocks model the conversion of jet kinetic
energy to $\g$ rays takes place at a radius $R\approx 2\Gamma^2 c t_v
\sim 2.7\cdot 10^{13}\,(\Gamma/300)^2 (t_{v}/10^{-2}~\rm s)$~cm, which
can vary widely depending on the jet bulk Lorentz factor $\Gamma$ and
the $\g$ ray flux variability time scale $t_v$.  The jet kinetic
energy is typically estimated from the observed isotropic-equivalent
$\g$-ray luminosity $L_\g$ and assuming that a fraction $\epsilon_e$
of the kinetic energy is converted to relativistic electrons which
promptly radiate most of their energy to $\g$ rays.  Random magnetic
field in the GRB jet is believed to arise when a fraction $\epsilon_B$
of the jet kinetic energy is converted to the magnetic field energy in
the shocks (see e.g. Ref.~\cite{Medvedev:1999tu,Gruzinov99}).  An
average value of the magnetic field and electron density can be
estimated, in the jet comoving frame, as $B \sim 5\cdot 10^4\,
(\epsilon_B/\epsilon_e)^{1/2} (L_\g/10^{52}~{\rm erg}~{\rm
  s}^{-1})^{1/2} (\Gamma/300)^{-3} (t_{v}/10^{-2}~\rm s)^{-1}$~G and
$n_e \sim 2\cdot 10^8\, \epsilon_e^{-1} (L_\g/10^{52}~{\rm erg}~{\rm
  s}^{-1}) (\Gamma/300)^{-6} (t_{v}/10^{-2}~\rm s)^{-2}$~cm$^{-3}$
(see e.g. Ref.~\cite{Razzaque:2004cx}).

Strong magnetic field from the central engine can also be present in
the GRB jet.  The toroidal component of the magnetic field of a
magnetar with surface magnetic field $B_0$ at $R_0\approx 10^6$~cm
drops to a value $B=B_0(R_0/R) \approx 10^8(B_0/10^{15}~{\rm
  G})(R/10^{13}~{\rm cm})^{-1}$~G at a dissipation radius $R$.  The
magnetic field from the central engine is globally ordered in the
emission region.  The coherence length scale of the random magnetic
field can be as small as the plasma skin
depth~\cite{Medvedev:1999tu,Gruzinov99}, however efficient conversion
of the shock energy to $\g$ rays requires a length scale of the order
of the comoving width of the plasma shell $\langle \Delta R \rangle
\approx \Gamma ct_v \sim 9\cdot 10^{10}\, (\Gamma/300)
(t_{v}/10^{-2}~\rm s)$~cm.  Because of relativistic beaming, only an
angular size scale $1/\Gamma$ of the jet surface is viewable.  Note
that, this also corresponds to a maximum length scale $\langle \Delta
R \rangle$ over which the random magnetic field can be fully ordered
due to causality~\cite{Gruzinov99}.  The jet half-opening angle
$\theta_{\rm jet}$ is much larger than $1/\Gamma$ during the prompt
$\g$-ray emitting phase.  Both the ordered and random magnetic fields
are mostly perpendicular to the jet axis, which is assumed along the
$z$ direction.

Synchrotron radiation from the visible patch of the jet surface can
reach the maximum polarization degree, $\approx 50\%$--$70\%$, if the
magnetic field is fully ordered in the patch and $\Gamma\theta_{\rm
  jet} \gg 1$.  Intrinsic curvature of the field, for example in case
of toroidal field configuration, in a large visible patch can reduce
the maximum polarization degree to $\approx 40\%$~\cite{Granot03,
  toma}.  Smaller scale random magnetic field, if dominant, can also
reduce the net polarization degree~\cite{Gruzinov99, toma}.  We
explore both the ordered and random field scenarios to calculate
photon-ALP mixing in the GRB jet.  Moreover, the emission region and
propagation region of the photons can be separated with different
magnetic field strengths and orientations (i.e. $\xi \ne 0$).  Faraday
rotation of the polarization plane can be important for synchrotron
radiation \cite{Rybicki:1979} only for a substantial magnetic field
component parallel to the beam direction (along the $z$ axis) and
below the optical frequencies, both situations are outside the scope
of this paper.  Mixing of the $A_\parallel$ and $A_\perp$ components
in our scenario takes place through the off-diagonal terms in the
mixing matrix [Eq.~(\ref{eq:evolxi})], due to $\xi$.  Additional
ordered magnetic field (e.g. in the wind of the progenitor star)
surrounding the GRB jet~\cite{Granot03}, if present and is
sufficiently strong, can modify some of the polarization effect that
we explore here.  However we ignore that for simplicity.

The two photon polarization components in synchrotron radiation can be
written in terms of the Bessel functions (see
e.g. Ref.~\cite{Rybicki:1979}) as\footnote{Note that
Ref.~\cite{Jackson:1998} uses exactly the opposite convention for the
polarization components.}
\ba
A_{\parallel} (\w) &=& \frac{\sqrt{3}\g_e^2\theta_e}{\w_c}
\sqrt{1+\g_e^2\theta_e^2} \,K_{1/3} \left(\frac{\w}{2\w_c}\right)~,
\nonumber \\
A_{\bot} (\w) &=& i \frac{\sqrt{3}\g_e}{\w_c}
(1+\g_e^2\theta_e^2) \,K_{2/3} \left(\frac{\w}{2\w_c}\right) \,,
\label{eq:sync_fields}
\ea
from a single electron with Lorentz factor $\g_e$ gyrating in the
$\bfB$ field. Here $\theta_e$ is the angle between the line of sight
and the plane containing the electron trajectory.  The characteristic
synchrotron frequency, in case $\theta_e\to 0$, is given by
\be
\w_c = \frac{3}{2} \frac{B\sin\eta}{B_{\rm crit}} \g_e^2 m_e\,,
\label{sync_wc}
\ee
where $\eta$ is the pitch angle between the electron's velocity and
$\bfB$.  The intensity of synchrotron radiation is given by
\be
\frac{d^2 I}{d\w d\Omega} = \frac{e^2\w^2}{4\pi^2} 
\left( |A_\parallel (\w)|^2 + |A_\bot (\w)|^2 \right)\,,
\label{sync_intensity}
\ee
and the emitted radiation is concentrated in a solid angle $d\Omega =
2\pi\sin\eta \, d\theta_e$.  The power emitted per unit frequency is
calculated by dividing the intensity with the orbital period of the
charge, $T= 2\pi\g_e m_e/eB$, after integrating over the solid angle
as
\be
P(\w) =  \frac{e^3\w^2 B\sin\eta}{4\pi^2 \g_e m_e} 
\int \left( |A_\parallel (\w)|^2 + 
|A_\bot (\w)|^2 \right) d\theta_e \,.
\label{sync_power1}
\ee
The degree of linear polarization for a mono-energetic electron is
given by~\cite{Rybicki:1979}
\be
\Pi_L \equiv \frac{P_\bot (\w) -P_\parallel (\w) }{P_\bot(\w)
+P_\parallel (\w)}~,
\label{eq:polz}
\ee
where $P_\bot(\w)$ and $P_\parallel(\w)$ are the powers emitted per
unit frequency in directions parallel and perpendicular to the
magnetic field, and can be calculated from Eq.~(\ref{sync_power1}).

The total synchrotron power from a distribution of
electrons\footnote{See e.g. Ref.~\cite{Rybicki:1979} for power-law
  distribution of electron Lorentz factor.} can be calculated by
performing the convolution of the power from each electron and by
integrating over $\g_e$.  In the keV--MeV range of our interest,
however, $\g$ rays from GRBs are modeled as synchrotron radiation from
the shock-accelerated electrons of a minimum Lorentz factor
$\g_{e,m}$.  The observed peak photon energy in the $E^2(dN/dE)$
energy spectrum (often denoted as $EF(E)$ or $\nu F_\nu$) corresponds
to the characteristic photon energy in Eq.~(\ref{sync_wc}), after
multiplying by a $\Gamma/(1+z)$ factor, as $E_{\rm pk} \sim
3.5\,(1+z)^{-1} (B\sin\eta/10^6~{\rm G})(\g_{e,m}/10^3)^2
\Gamma_{300}$~MeV.  The typical GRB redshift is $z\approx 1$--2.
Higher energy photons, but not too far above $E_{\rm pk}$, can be
modeled as synchrotron radiation from a power-law distribution of
electrons above $\g_{e,m}$ and do not couple to ALPs in our present
study.

To explore photon-ALP mixing in the GRB jet environment, we solve the
field evolution equation [Eq.~(\ref{eq:evolxi})] with mixing matrix
elements [Eq.~(\ref{matrix_elements})] derived from GRB environment
parameters, and with initial electromagnetic field input from
Eq.~(\ref{eq:sync_fields}). Note that the comoving frame values for
the GRB parameters are used to  evaluate photon-ALP mixing, and the
resulting effect show up in the comoving frame frequency $\omega$.
The observed photon energy is $E=\omega \Gamma/(1+z)$.  We
calculate the effect of photon-ALP mixing on the polarization pattern
by using $A_\perp$ and $A_\parallel$ from solutions of the evolution
equation [Eq.~(\ref{eq:evolxi})] to find the linear degree of
polarization as
\be
\Pi_{L,\rm ALP} \equiv \frac{P_{\bot,\rm ALP} (\w) -P_{\parallel_,\rm
ALP} (\w) }{P_{\bot, \rm ALP}(\w) +P_{\parallel, \rm ALP} (\w)}~,
\label{eq:syncr}
\ee
and compare with Eq.~(\ref{eq:polz}), without photon-ALP mixing.  We
also define a flux modification factor, from Eq.~(\ref{sync_power1}),
as
\be
\rho = P(\w)_{\rm ALP} / P(\w)~,
\label{eq:ratio}
\ee
which shows any deviation from the synchrotron spectra due to
photon-ALP mixing in the GRB jet.  We discuss results from our
investigation next.

\section{Results and Discussion}

For the nominal values of the GRB parameters $B_T = 10^6$~G, $n_e =
10^8$~cm$^{-3}$, $L=10^{11}$~cm, and for the photon-ALP coupling
constant $g_{a\g} = 8.8\cdot 10^{-11}$~GeV$^{-1}$ which is very close
to the current CAST limit~\cite{cast}; strong mixing of photons and
ALPs takes place in the GRB jet when $m_a \le \sqrt{2g_{a\g} \omega
  B_T} \lesssim 10^{-6}\sqrt{\omega/{\rm keV}}$~eV, from the condition
$\Delta_a^2 \le 4\Delta_{a\g}^2$.  Indeed the photon-ALP mixing term
$\Delta_{a\g}$ dominates other terms [Eq.~(\ref{matrix_elements})] in
the mixing matrix for the nominal GRB parameters, and $\Delta_{\rm
  osc} \approx 2\Delta_{a\g} \sim L^{-1}$ [Eq.~(\ref{eq:probosc})].
The mixing angle $\theta$ is also maximized in this case, as
$(\Delta_\parallel - \Delta_a) < \Delta_{a\g}$.  The off-diagonal
rotation term $\propto (\Delta_\parallel - \Delta_\perp) =
(3/2)\Delta_{\rm QED} \sim 2\cdot 10^{-12} (\omega/{\rm
  keV})$~cm$^{-1}$ is small at low $\omega$ for the nominal GRB
parameters, but can become significant at high $\omega$.  Thus it is
important to keep all terms in the mixing matrix and solve the
evolution equation [Eq.~(\ref{eq:evol})] numerically with
frequency-dependent initial conditions from
Eq.~(\ref{eq:sync_fields}).  Photon-ALP conversion mostly takes place
in a broad observed energy range of $E \approx
(12$--$560)(\Gamma/100)(1+z)^{-1}$~keV [Eq.~(\ref{LHenergy})] for our
reference parameters.

\FIGURE{\epsfig{file=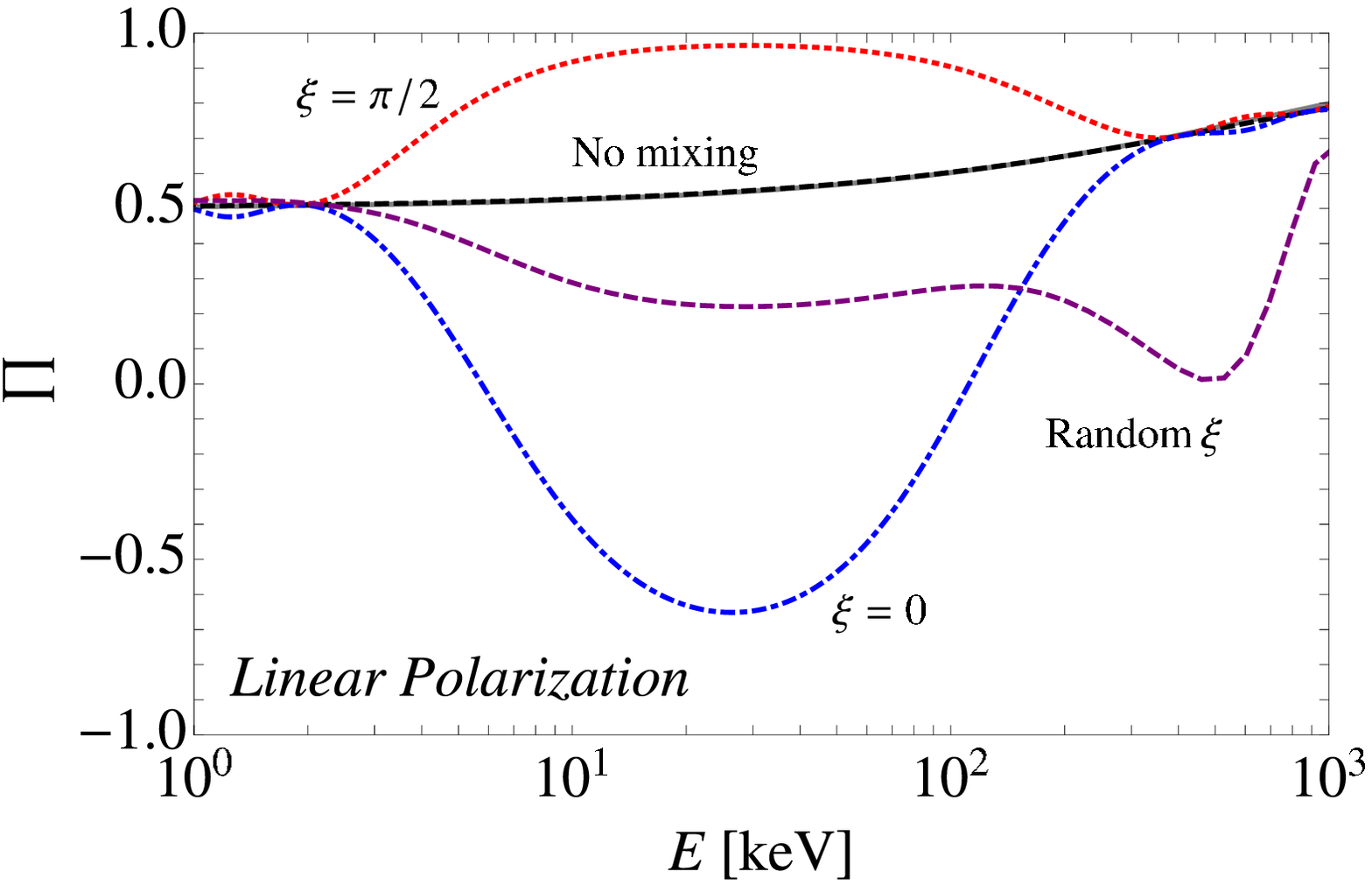, height=7cm} \caption{Linear
    photon polarization with and without ALP mixing in the GRB jet for
    the nominal GRB parameters $B_T = 10^6$~G, $n_e = 10^8$~cm$^{-3}$
    and $L=10^{11}$~cm.  We used a photon-ALP coupling parameter value
    $g_{a\g} = 8.8\cdot 10^{-11}$~GeV$^{-1}$ along with ALP mass $m_a
    = 10^{-7}$~eV.  The GRB is assumed to be at redshift $z=2$ with a
    jet bulk Lorentz factor $\Gamma = 100$.  Also the synchrotron
    emission from the GRB is assumed to peak at $\approx 660$~keV in
    the observer's frame.  The polarization degree without photon-ALP
    mixing is shown as the black dashed line obtained by solving the
    evolution equation [Eq.~(\ref{eq:evolxi})].  The solid gray line
    is the expected polarization from synchrotron theory.  The role of
    the final $A_\parallel$ and $A_\perp$ are interchanged from the
    initial configuration while $\xi$ changes from 0 (blue dot-dashed
    line) to $\pi/2$ (red dotted line).  Total polarization from many
    identical domains but with random $\xi$ is also shown (purple
    dashed line).}
\label{fig:grb_polz}}

Figure~\ref{fig:grb_polz} shows the effects of photon-ALP mixing in
the GRB jet with nominal parameters as mentioned above with $\Gamma =
100$ and $z=2$.  The peak of the synchrotron radiation is assumed at
$\omega_c = 20$~keV in the comoving GRB jet frame (2~MeV in the rest
frame of the source or $\approx 660$~keV in the observer's frame).
The initial polarization obtained by numerically solving the evolution
equation [Eq.~(\ref{eq:evolxi})], without photon-ALP mixing, is
plotted with the black dashed line, which agrees with theoretical
expectation (solid gray line).  The results for photon-ALP mixing are
plotted for two cases, $\xi=0$ (blue dot-dashed line) and $\xi=\pi/2$
(red dotted line).  The change in polarization from the $\xi=0$ case
to the $\xi=\pi/2$ case can be understood as the magnetic field
orientation in the initial production region and propagation region
being aligned parallel with each other in the former case and being
aligned perpendicular to each other in the latter case.  In other
words, as an inspection of the mixing matrix in Eq.~(\ref{eq:evolxi})
reveals, the $A_\parallel$ and $A_\perp$ in the final states are
interchanged from the initial configuration for $\xi=\pi/2$.
Observations in limited energy bands, however, can not distinguish
between the two extreme cases and is expected to be intermediate,
since polarimeters measure the absolute degree of polarization, highly
correlated with $\xi$ within a unique energy band. On the other hand,
a change in polarization degree in different energy bands, different
from the synchrotron radiation pattern, can be used to search for
photon-ALP mixing signature.

Time-resolved measurements over small intervals and around the pulses
in the GRB light curves are important to ensure that emission from
only a small bright spot, in which we assume the magnetic field to be
fully coherent, of the jet surface contributes in each case.  For
longer exposure, contributions from many domains (assumed identical)
on the jet surface can contribute.  This case, where each domain is
assumed to have completely ordered field within and only the
orientation of the magnetic field direction $\xi$ is assumed random,
is also shown in Fig.~\ref{fig:grb_polz} with the purple dashed line.

\FIGURE{\epsfig{file=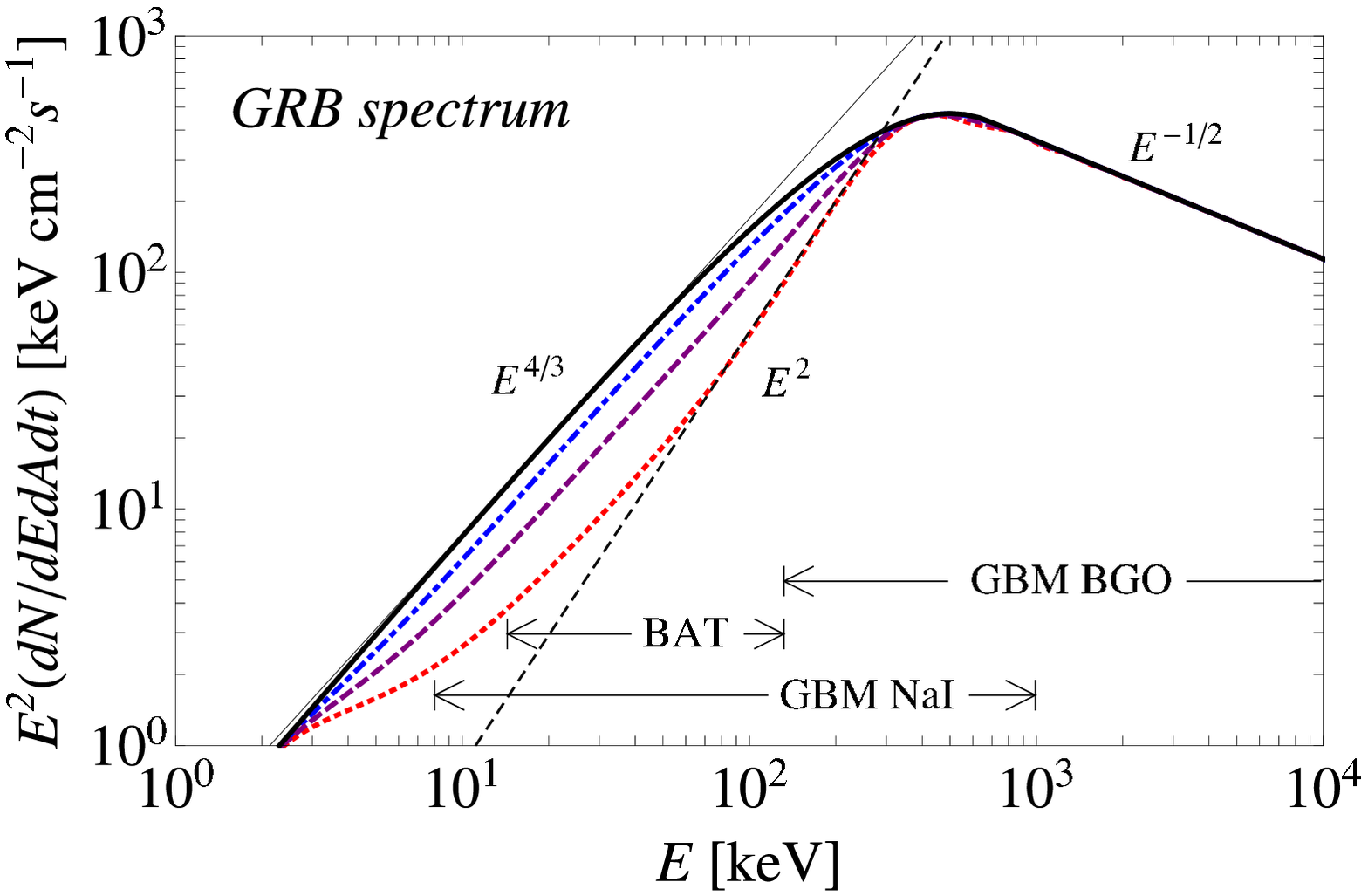, height=7cm} \caption{ Phenomenological
    GRB energy spectrum with and without photon-ALP mixing with the
    same parameters used in Fig.~\ref{fig:grb_polz}.  The spectrum for
    no photon-ALP mixing (solid thick black line) is plotted using the
    Band spectrum with peak photon energy $E_{\rm pk,Band} = 500$~keV,
    and low (high) energy power-law slope $\alpha_{\rm Band}\approx
    -0.6$ ($\beta_{\rm Band} = -5/2$).  Asymptotically the low energy
    power-law slope coincides with the expected spectrum with
    $\alpha_\g = -2/3$ from synchrotron theory (thin solid gray line).
    The effects of photon-ALP mixing are plotted by multiplying the
    Band spectrum with the suppression factor in Eq.~(\ref{eq:ratio})
    for the $\xi=0$ (blue dot-dashed line), $\xi=\pi/2$ (red dotted
    line) and random $\xi$ (purple dashed line) cases.  As can be
    seen, the observed spectra for the $\xi=\pi/2$ case can be steeper
    than the synchrotron spectrum in a limited energy range (thin
    dashed gray line).  Also plotted are the energy bands in which
    {\em Swift} BAT and {\em Fermi} GBM instruments are
    sensitive.}\label{fig:grb_spectrum}}

Figure~\ref{fig:grb_spectrum} shows the effects of photon-ALP mixing
on the GRB spectrum.  A phenomenological photon spectrum ($dN/dE$),
called the Band spectrum~\cite{Band93}, is plotted (thick black curve)
with the low-energy index $\alpha_{\rm Band} \approx -0.6$, high
energy power-law index $\beta_{\rm Band}=-5/2$ and a peak energy
$E_{\rm pk, Band} = 500$~keV.  The GRB is assumed to be at $z=2$ as in
Fig.~\ref{fig:grb_polz} with all other parameters for photon-ALP
mixing the same as those used for Fig.~\ref{fig:grb_polz}.  We assume
that synchrotron radiation from minimum energy electrons dominates
below the start of the high energy power-law part of the spectrum at
$(2+\alpha_{\rm Band})E_{\rm pk, Band} \approx 700$~keV, similar to
$\omega_c$ in the jet comoving frame.  The effective low energy
power-law index~\cite{Preece:1998jy}, corresponding to the synchrotron
theory, is then $\alpha_\g \approx -2/3$ ($E^{4/3}$ in the $E^2 dN/dE$
or $\nu F_\nu$ spectrum) as plotted with a thin solid gray line.
Approximately $20\%$ of the GRB spectra are steeper than this
``synchrotron death line'', and in $\approx 5\%$ of the time-resolved
spectra of bright GRBs the spectral deviation is statistically
significant~\cite{Kaneko:2006qe}.  As shown in
Fig.~\ref{fig:grb_spectrum} photon-ALP mixing for $\xi=\pi/2$ case
(red dotted line) can change the low-energy spectrum to as steep as
$dN/dE\propto E^0$ (thin gray dashed line) from the synchrotron model,
depending on the parameters we used.  The change in the spectrum is
not as significant, however, for the $\xi=0$ and random $\xi$ cases.

The $\xi=\pi/2$ case should be less frequent in nature as evidenced by
the fraction of GRB spectra that violates the ``synchrotron death
line''.  High polarization degree, up to 100\%, is expected in these
cases (Fig.~\ref{fig:grb_polz}).  Indeed the peak-resolved spectra of
GRB 021206 with $(80\pm 20)\%$ polarization~\cite{detect} show
low-energy index as hard as $\alpha_{\rm Band} = -0.42\pm
0.05$~\cite{Wigger:2007yf}.  A larger sample of GRBs with correlated
high polarization and steep low-energy spectrum detected with future
polarimeters will be instrumental to probe the photon-ALP mixing in
GRB jets.  Other explanation of steep spectrum by black body, jitter
radiation, inverse Compton scattering etc.\ (see
e.g. Ref.~\cite{Ghirlanda03}) do not generally change the polarization
pattern the way photon-ALP mixing does and as we discussed here.

\FIGURE{\epsfig{file=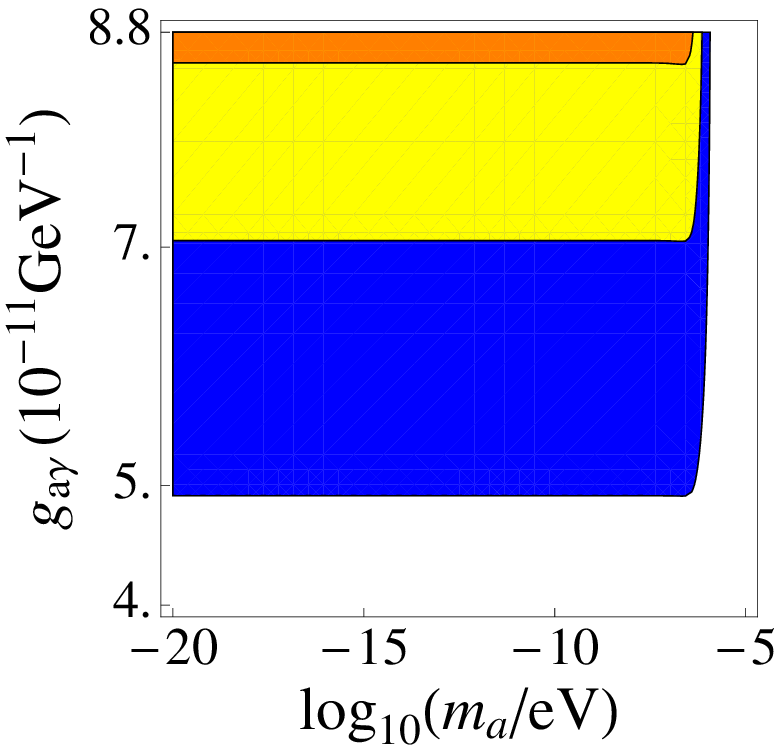, height=7cm}\epsfig{file=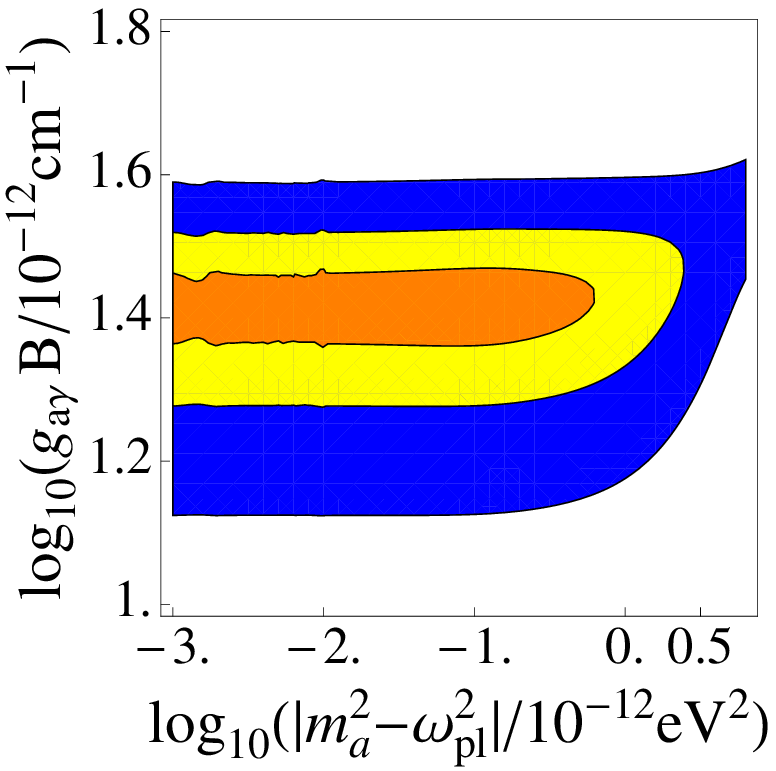,
    height=7cm} \caption{Contour plots of the flux suppression
    factor in the ALP parameter plane with fixed GRB parameters (left
    panel), and in the mixed ALP-GRB parameter plane (right panel).
    The outer, middle and inner contours depict the regions in which
    the flux suppression factor, see Eq.~(\ref{eq:ratio}), reaches
    70\%, 50\% and 40\%, respectively.  For $m_a < 10^{-6}$~eV, the
    effect becomes independent of the ALP mass and is restricted to a
    rather narrow range of $g_{a\g}B$ values.}
\label{fig:contour}}

Figure~\ref{fig:contour}, left panel, depicts the flux suppression
factor $\rho$, see Eq.~(\ref{eq:ratio}), in the $m_a$--$g_{a\g}$ plane
with fixed GRB parameters ($B_T = 10^6$~G, $n_e = 10^8$~cm$^{-3}$ and
$L=10^{11}$~cm).  The outer, middle and inner contours depict the
regions in which the flux suppression factor reaches 70\%, 50\% and
40\%, respectively.\footnote{A smaller suppression factor corresponds
  to a larger conversion probability of photons to ALPs.}  The
suppression effect has been averaged in the (10--500) keV energy
window. A flux suppression factor $<40\%$, requires a photon ALP
coupling parameter close to the current CAST limit.  As previously
discussed, the mixing angle is maximized when the photon-ALP mixing
term dominates the system evolution, and, for sufficiently small ALP
masses ($m_a< 10^{-6}$~eV), the effect becomes independent of the ALP
mass.  In the pure axion cold dark matter (CDM) scenario, if the PQ
symmetry is restored after inflation, a lower mass bound
$m_a>10^{-5}$~eV applies in order to not overclose the universe (see
Ref.~\cite{pdg} and references therein).  However, if inflation takes
place after the PQ transition, much smaller values for the CDM axion
mass are still allowed, see Refs.~\cite{Linde:1987bx,
  Tegmark:2005dy,Arvanitaki:2009fg,Visinelli:2009zm}. The ALP case is
more complicated, since these particles constitute a dark matter
candidate only under certain conditions.  For instance, if ALPs couple
exclusively to photons, they are excluded as CDM
candidates~\cite{Masso:2006id}. Consequently, no lower ALP mass bound
is shown in Fig.~\ref{fig:contour} (left panel), since the role of
ALPs as CDM particles depends highly on the underlying theoretical
model.  

Figure~\ref{fig:contour}, right panel, depicts the flux suppression
factor $\rho$ same as in the left panel.  The contours are plotted in
the $g_{a\g}B$ and $|m_a^2 - \omega_{\rm pl}^2|$ plane, where both the
quantities are closely related to the wave numbers
[Eq.~(\ref{matrix_elements})].  Comparing the ranges of $g_{a\g}$ and
$g_{a\g}B$ values from the plots, significant ($\lesssim 70\%$) flux
suppression takes place for $B \sim 4\cdot 10^5$--$3\cdot 10^6$~G in
the observed keV--MeV range.  Thus detection of photon-ALP mixing
effect in GRB data can, in principle, be used to probe the magnetic
field value in the GRB jet, which is somewhat uncertain.  Note that a
much higher, $\sim 10^9$~G, field with a $\sim 10^6$~cm coherence
length, corresponding to the neutron star radius used in
Refs.~\cite{rubbia,bassan} gives no photon-ALP mixing effect in the
GRB jet.  However, the radius of $\g$-ray emission region is likely to
be large to avoid $e^+e^-$ pair creations by the photons and
thermalization.  The magnetic field in the jet is thus likely to be
small, typical to the values that we used, at this large radius.

Mixing of photons with ALPs, for propagation in the IGMF with
generally assumed magnetic field $B_{\rm IGMF} = 1$~nG and particle
density $n_e = 10^{-7}$~cm$^{-3}$, takes place in the frequency range
[see Eq.~(\ref{LHenergy})] $\omega_L \approx 10^7 (m_a/10^{-7}~{\rm
  eV})^2 (B_{\rm IGMF}/{\rm nG})^{-1}$~GeV and $\omega_H \approx
6\cdot 10^9(B_{\rm IGMF}/{\rm nG})^{-1}$~GeV for the same $g_{a\g}$
parameter from the CAST limit.  The contribution of plasma frequency
to $\omega_L$ becomes dominant for ALP mass $m_a \ll 10^{-14}
\sqrt{n_e/10^{-7}~{\rm cm}^{-3}}$~eV from the condition $\omega_{\rm
  pl} \gg m_a$ in Eq.~(\ref{LHenergy}).  The corresponding $\omega_L
\approx 0.1 (n_e/10^{-7}~{\rm cm}^{-3}) (B_{\rm IGMF}/{\rm
  nG})^{-1}$~keV becomes constant.  The oscillation wave number is
$\Delta_{\rm osc} \approx \Delta_{a\g}$ in this asymptotic range, and
the oscillation probability [Eq.~(\ref{eq:probosc})] is $P_{a\g}
\approx (\Delta_{a\g} L)^2 \approx 2\cdot 10^{-3} (B_{\rm IGMF}/{\rm
  nG})^2 (L/{\rm Mpc})^2$ for Mpc scale coherence length.  Thus
photon-ALP mixing in the IGMF can be important over Gpc scale source
distance and wash-out the source signature only if the IGMF is of the
order of nG and the ALP mass is smaller than $10^{-14}$~eV.  This
result is compatible with mixing effect in the IGMF for ultra-light
ALPs explored in Ref.~\cite{bassan}.  In fact these two mixing
scenarios, in-source and in the IGMF, are complementary to each other
and cover a huge range of ALP mass.  Detection of source signatures can
be used to constrain the ALP mass as well as to put limit on the IGMF.
Indeed, there are hints from recent studies of ultra high-energy
cosmic ray data and TeV blazars that the IGMF can be much smaller than
a nG~\cite{IGMF}, in which case the polarization and spectral
signatures of in-source photon-ALP mixing that we explored will not be
destroyed.  Photon-ALP mixing in the $\sim \mu$G galactic magnetic
field over kpc coherence length scale is also negligible.

\section{Conclusions}

Axions and axion-like particles appear in many extentions of the
standard model of particle physics.  Photon-axion/ALP mixing in the
presence of an external electromagnetic field constitutes one of the
most exploited signals for astrophysical and laboratory axion and ALP
searches.  Gamma-ray bursts are the most powerful source of keV--MeV
photons in nature, which are most probably synchrotron radiation.
These photons originate and propagate inside the GRB jet with high
magnetic field.  We have shown that strong photon-ALP conversion takes
place in GRB jet in the $\sim 100$~keV observed energy range,
distorting the standard synchrotron polarization pattern.  We have
also shown that when the magnetic field direction in the photon
propagation coherence length is perpendicular to the magnetic field
direction in the synchrotron radiating region, the photon energy
spectrum will be steeper than the expected spectrum from synchrotron
theory, thus providing an explanation for the anomalous spectra of
$\sim 20\%$ of the observed GRBs.  We found that the photon-ALP
conversion occurs within a large range of possible GRB and ALP
parameters, being almost independent of the ALP mass for sufficiently
small ALP masses ($m_a<10^{-6}$~eV).  Further modification due to
mixing in the intergalactic magnetic field is not expected in case the
IGMF is $\lesssim 1$~nG and/or the ALP mass is $\gtrsim 10^{-14}$~eV.

Large statistics expected to be collected by a number of future
missions that are devoted to measure GRB polarization in the keV--MeV
range will be crucial to search for ALP signals due to their mixing
with photons inside GRBs.

\section{Acknowledgments}

We thank Mikhail Medvedev, Carlos Pe\~{n}a-Garay, John Ralston and
Kenji Toma for helpful discussions, and Justin Finke for useful
comments on the manuscript.  Work of O.~M. was supported by the MICINN
(Spain) Ram\'on y Cajal contract, AYA2008-03531 and CSD2007-00060.
Work of S.~R. was performed at and under the sponsorship of the Naval
Research Laboratory (USA) and was partially supported by the Fermi
Cycle 2 Guest Investigator Program of NASA (USA).

\end{document}